\documentclass[11pt]{article}
\usepackage{verbatim}
\usepackage{fullpage}
\usepackage{amsfonts}
\usepackage{amssymb}
\usepackage{color}
\usepackage{graphicx}
\usepackage{helvet}
\usepackage{latexsym}
\usepackage{amsgen}
\usepackage{amsmath}
\usepackage{amsopn}
\usepackage{amsxtra}

\newtheorem{theorem}{Theorem}[section]
\newtheorem{lemma}[theorem]{Lemma}
\newtheorem{proposition}[theorem]{Proposition}

\newtheorem{corollary}[theorem]{Corollary}

\newenvironment{proof}[1][Proof:]{\begin{trivlist}
\item[\hskip \labelsep {\bfseries #1}]}{\end{trivlist}}

\newcommand{\qed}{\hfill\rule{2mm}{2mm}}

\newcommand{\ket}[1]{|#1\rangle}
\newcommand{\braket}[2]{\langle #1|#2\rangle}
\newcommand{\ketbra}[2]{|#1\rangle\langle #2|}

\renewcommand{\S}{\mathcal{S}}
\newcommand{\Z}{\mathbb{Z}}

\newcommand{\I}{\mathcal{I}}
\newcommand{\J}{\mathcal{J}}

\newcommand{\mix}{\operatorname{mix}}

\begin{document}

\title{Quantum speedup of classical mixing processes\thanks{This
material is based upon work supported by the National Science
Foundation under Grant No. 0523866.  Part of this research was done
while the author was visiting LRI, Universit\'{e} Paris-Sud, Orsay, France.}}
\author{Peter C. Richter\thanks{Department of Computer Science,
Rutgers University, Piscataway, NJ 08854.  {\tt richterp@cs.rutgers.edu}}}
\date{}
\maketitle

\begin{abstract}
Most approximation algorithms for \#P-complete problems (e.g.,
evaluating the permanent of a matrix or the volume of a polytope) work
by reduction to the problem of approximate sampling
from a distribution $\pi$ over a large set $\S$.  This problem is
solved using the {\em Markov chain Monte Carlo} method: a sparse,
reversible Markov chain $P$ on $\S$ with stationary distribution $\pi$
is run to near equilibrium.  The running time of this random walk
algorithm, the so-called {\em mixing time} of $P$, is $O(\delta^{-1} \log
1/\pi_*)$ as shown by Aldous, where $\delta$ is the spectral gap
of $P$ and $\pi_*$ is the minimum value of $\pi$.  A natural question
is whether a speedup of this classical method to $O(\sqrt{\delta^{-1}}
\log 1/\pi_*)$, the diameter of the graph underlying $P$, is possible
using {\em quantum walks}.

We provide evidence for this possibility using quantum walks that {\em
decohere} under repeated randomized measurements.  We show: (a)
decoherent quantum walks always mix, just like their classical
counterparts, (b) the mixing time is a robust quantity, essentially
invariant under any smooth form of decoherence, and (c) the mixing
time of the decoherent quantum walk on a periodic lattice $\Z_n^d$ is
$O(n d \log d)$, which is indeed $O(\sqrt{\delta^{-1}} \log 1/\pi_*)$ and
is asymptotically no worse than the diameter of $\Z_n^d$ (the obvious
lower bound) up to at most a logarithmic factor.
\end{abstract}

%\thispagestyle{empty}
%\newpage
%\setcounter{page}{1}

\section{Introduction}

\subsection{Markov chain Monte Carlo and quantum walks}

A rich theory has been developed for computing approximate solutions
to problems in combinatorial enumeration and statistical physics which
are \#P-complete and therefore unlikely to have efficiently
computable exact solutions.  Among the highlights
are randomized polynomial-time approximation schemes for evaluating
the permanent of a nonnegative matrix (Jerrum, Sinclair, and Vigoda
\cite{JSV}), the volume of a convex polytope (Dyer, Frieze, and Kannan
\cite{DFK}), and the partition functions of monomer-dimer and
ferromagnetic Ising systems (Jerrum and Sinclair \cite{JS1,JS2}).
At the heart of these algorithms (and consuming most of the running
time) is a subroutine for approximate sampling from a particular
distribution $\pi$ over a large set $\S$.  This problem is solved
using the {\em Markov chain Monte Carlo} (MCMC) method: a sparse,
reversible {\em Markov chain} (stochastic matrix) $P$ on $\S$ with
stationary distribution $\pi$ is run from an arbitrary initial state
to a random state distributed very close to $\pi$.  The time required
to guarantee this convergence, or {\em mixing}, is the so-called {\em
mixing time} $\tau_{\mix}$. Bounding $\tau_{\mix}$ is typically the
chief technical hurdle in analyzing the running time of the overall
algorithm.  It is reduced to the problem of estimating the spectral
gap $\delta$ of $P$ by Aldous' inequality \cite{Ald}
\begin{equation}\label{aldous}
\delta^{-1} \leq \tau_{\mix} \leq \delta^{-1} \: (1 + \frac{1}{2} \log 1/\pi_*)
\end{equation}
where $\pi_*$ is the minimum value of $\pi$.  Since $\delta$ is
generally no easier to estimate directly than $\tau_{\mix}$, we often
bound the {\em conductance} $\Phi$ (a geometric parameter of the
chain) and appeal to the Cheeger inequality (see e.g., Alon \cite{Alo}):
\begin{equation}
\frac{1}{2} \Phi^2 \leq \delta \leq 2\Phi
\end{equation}
Further inequalities (e.g., involving congestion of multicommodity
flows through the chain) are often invoked on top of these.

Notice that Aldous' inequality (\ref{aldous}) is tight with respect to
the spectral gap.  However, it is also somewhat unsatisfactory in that
although the diameter of the graph underlying $P$ (the obvious lower
bound for sampling from its stationary distribution) scales like
$O(\sqrt{\delta^{-1}} \log 1/\pi_*)$, the dependence is $\delta^{-1}$
in Aldous' inequality due to the Gaussian-like spreading behavior of
random walks.  In MCMC applications, the additional factor of
$\sqrt{\delta^{-1}}$ can mean the
difference between, say, an $n^3$ and $n^5$ algorithm!  Removing it
would imply a considerable improvement in both the {\em known} upper
bounds (since existing estimates of $\delta$ could be used in
conjunction with a sharper inequality than Aldous') and the {\em true}
upper bounds (since Aldous' inequality is tight and thus a sharper
inequality could only come from a faster approximate sampling method).
Thus, a natural question is whether there is a way to modify the standard 
MCMC method to obtain a speedup to $O(\sqrt{\delta^{-1}} \log 1/\pi_*)$.
%\footnote{Formally, is there a way to abstract the generic
%MCMC method $P$ and devise a more clever route to outputting a
%near-uniform sample, while still being able to say that ``whenever we
%can perform this MCMC method $P$, we can instead perform this better
%method''.  More precisely, we can at each vertex $x$ efficiently obtain the
%transition probabilities $p_{yx}$, apply to them a fixed function $f$
%to obtain values $q_{yx}$ (ie, depending on their values, but not $x$
%itself.. the procedure must be local/oblivious; also, if $f$ must map
%a zero transition coefficient to zero), then transit to an adjacent
%vertex according to the probability/amplitude $q_{yx}$.  For example,
%the standard MCMC method has $f=Id$.}
This seems unlikely using classical randomized methods: Chen, Lovasz,
and Pak \cite{CLP} have shown that we can sometimes speed up mixing by
{\em lifting} a Markov chain, but this requires both knowledge of the
chain's global structure and its use in solving an NP-hard flow
problem to find low-congestion paths along which to ``route'' probability mass
efficiently.  However, although lifting the chain seems unlikely to be
practical, an idea that might work is {\em quantizing} the chain.

Why might a quantized Markov chain, or {\em quantum walk}, help us
reach $\pi$ more quickly?  Two reasons are: (1) a quantum walk is as
simple to realize as its classical counterpart (i.e., it is computable
locally and online, unlike a classical lifting of the chain), and (2)
there is empirical (and some theoretical) evidence that quantum walks
tend to propagate and ``spread'' probability mass across $\S$ in time
on the order of the diameter on precisely the same low-dimensional
graphs that trip up their classical counterparts.  Based on these
observations, the possibility of obtaining a quantum speedup
for the mixing problem has been pursued by Nayak et
al. \cite{NV,ABNVW}, Aharonov et al. \cite{AAKV}, Moore and Russell
\cite{MR}, Gerhardt and Watrous \cite{GW}, and Richter \cite{Ric}.

We remark that a quantum speedup theorem of the sort we seek has
already been proven for the {\em hitting} problem, in which we search
(rather than sample from) the states of a Markov chain:  Szegedy
\cite{Sze} proved a quadratic quantum speedup for the {\em hitting
time} of any symmetric Markov chain, generalizing considerably the
celebrated search algorithm of Grover \cite{Gro} and
implying quantum speedups for structured search problems such as
element distinctness (Ambainis \cite{Amb1}), triangle finding
(Magniez, Santha, and Szegedy \cite{MSS}), matrix product verification
(Burhman and Spalek \cite{BS}), and group commutativity testing
(Magniez and Nayak \cite{MN}).  It is this success which inspires us
to investigate the possibility of a quantum speedup for the mixing
problem with the goal of transferring the speedup to MCMC algorithms.

\subsection{Our contributions}

We present evidence of a possible quantum MCMC speedup to
$O(\sqrt{\delta^{-1}} \log 1/\pi_*)$ using quantum walks that {\em
decohere} under repeated randomized measurements.  Decoherence (in
small amounts) was first identified as a way to improve spreading and
mixing properties in numerical experiments performed by Kendon and
Tregenna \cite{KT} and in analytical estimates by Fedichkin et
al. \cite{FST,SF1,SF2}.  On the other hand, high rates of
decoherence in quantum walks have been shown to degrade mixing
properties substantially by the quantum Zeno effect (Alagic and
Russell \cite{AR}).  For an excellent survey of these and other
aspects of decoherent quantum walks, see Kendon \cite{Ken}.

Our technical contributions are as follows:  First, we show that for
any symmetric Markov chain $P$, we can generate an arbitrarily good
approximation to the uniform stationary distribution $\pi$ of $P$ by
subjecting the continuous-time quantum walk $U_{ct}(P) = \exp(iP)$ to
reasonably ``smooth'' decoherence.  Thus, decoherent
quantum walks (which are non-unitary) offer a way of circumventing an
obstacle first identified by Aharonov et al. \cite{AAKV}, who observed
that {\em unitary} quantum walks often converge (in the time-averaged
sense) to highly non-uniform distributions.

Second, we show that the optimal mixing time of a decoherent quantum
walk is a robust quantity, in that it remains essentially invariant
under any sufficiently smooth form of decoherence.  In particular,
decoherent quantum walks undergoing repeated Cesaro-averaged
\cite{AAKV,MR,GW} and Bernoulli/Poisson-averaged \cite{KT,AR}
measurements are nearly equivalent in mixing efficiency.  The proof
applies more generally to a game involving time-dependent Markov chains
(not necessarily describing quantum phenomena) and may be of
independent interest.

Third, we prove a theorem on threshold mixing of quantum walks on
(Cartesian) graph powers in order to show that the decoherent
continuous-time quantum walk $U_{ct}(P(G))$ on a periodic lattice
$G=\Z_n^d$ (where $P(G)$ denotes the standard Markov chain on $G$) can
be used to generate a good approximation to the uniform stationary
distribution $\pi$ of $P$ in time $O(n d \log d)$.  This upper bound is
asymptotically no worse than the diameter of $\Z_n^d$ (the obvious
lower bound) up to at most a logarithmic factor and is
$O(\sqrt{\delta^{-1}} \log 1/\pi_*)$ for both high-dimensional {\em and}
low-dimensional lattices (unlike its classical counterpart).  For
$d=1$, this proves a conjecture of Kendon
and Tregenna \cite{KT} based on numerical experiments.\footnote{In
fact this is not quite true: Kendon and Tregenna \cite{KT} conjectured
this for the discrete-time Hadamard walk.  We prove it for both
walks.}  For $d \geq 1$, it extends the results of Fedichkin et
al. \cite{FST,SF1,SF2} by confirming $O(n)$ and $O(d \log d)$ scaling
(suggested by their analytical estimates and numerical experiments
in regimes of both high and low decoherence) of the fastest-mixing
walk, which they conjectured to be decoherent rather than unitary.  We
briefly discuss the prospects for extending this result to the
discrete-time Grover walk $U_{dt}(P(\Z_n^d))$ \cite{IKK,TFMK,MBSS}.

Previously, mixing speedups had been proven only for the unitary
quantum walks of Nayak et al. \cite{NV,ABNVW} and Aharonov et
al. \cite{AAKV} (on the cycle) and of Moore and Russell \cite{MR} (on
the hypercube).  Thus, our work shows that introducing a small amount
of decoherence to a quantum walk can simultaneously force convergence
to the uniform distribution while preserving a quantum mixing speedup,
an advantageous combination for algorithmic applications.\footnote{Alagic
and Russell \cite{AR} exhibit a decoherent quantum walk on the
hypercube which converges to the uniform distribution and preserves
the quantum mixing speedup proven by Moore and Russell \cite{MR}, but
in contrast to our work, it is their single (carefully chosen) final
measurement that {\em forces} uniform convergence, not the decoherence
itself (which only adds ``noise'' that if small enough does not
destroy the speedup).}

\section{Preliminaries}

\subsection{Markov chains}

Let $P$ be a {\em Markov chain} (stochastic matrix) on the set
$\S$ ($|\S| = N$) which is {\em irreducible} (strongly connected); then
it has a unique distribution $\pi$ which is {\em stationary} (i.e., satisfies
$P \pi = \pi$).  Moreover, $\pi$ is strictly positive: $\pi_* := \min_x \pi_x$
satisfies $\pi_* > 0$.  In particular, if $P$ is symmetric, then $\pi$
is the uniform distribution $u$ (the $N$-dimensional column vector
with each component equal to $1/N$).  If $G$ is an undirected graph,
we denote the standard (``simple random walk'') Markov chain on $G$ by $P(G)$.

A Markov chain $P$ which is both irreducible and {\em aperiodic}
(non-bipartite) is by definition {\em ergodic} and satisfies
\begin{equation}
P^t \rightarrow \pi 1^\dagger = [\pi \cdots \pi] \textrm{ as } t \rightarrow
\infty
\end{equation}
where $1^\dagger$ is the $N$-dimensional row vector with
each component equal to one.  We can thus define the {\em (threshold)
mixing time}
\begin{equation}
\tau_{\mix} := \min \{ T : \frac{1}{2}||P^t - \pi 1^\dagger||_1 \leq
\frac{1}{2e} \: \forall t \geq T \}
\end{equation}
where $|| \cdot ||_1$ is the matrix 1-norm.  The mixing is {\em
perfect} if $P^t = \pi 1^\dagger$.  Let $\delta := 1 - ||P
|_{\pi^\perp}||_2 > 0$ be the {\em spectral gap} of $P$.  We say that
$P$ is {\em reversible} if the matrix $DPD^{-1}$ is symmetric, where
$D$ is the diagonal matrix $D(x,x) = \sqrt{\pi_x}$.  A precise
statement of Aldous' inequality (\ref{aldous}) is:
\begin{theorem}[Aldous \cite{Ald}]\label{ds-ineq}
Let $P$ be a reversible, ergodic Markov chain with stationary
distribution $\pi$ and spectral gap $\delta$.  Then its mixing
time satisfies $\delta^{-1} \leq \tau_{\mix} \leq \delta^{-1} \: (1 +
\frac{1}{2} \log 1/\pi_*)$.
\end{theorem}

We will also use the {\em maximum pairwise column distance}
$d(P) := \max_{x,x'} \frac{1}{2}||P(\cdot,x) - P(\cdot,x')||_1$
to estimate the mixing time.  It is related to the matrix 1-norm
distance by the inequality:
\begin{equation}
\frac{1}{2}||P - \pi 1^\dagger||_1 \leq d(P) \leq ||P - \pi 1^\dagger||_1
\end{equation}
%In light of this inequality and the next proposition, we sometimes
%consider the {\em threshold mixing time} of a Markov chain $P$ to be
%the first time in which $d(P^t)$ is bounded below one by a positive
%constant (essentially equivalent to the definition of $\tau_{\mix}$
%above.

The following propositions (see \cite{Ric} for proofs) can be used to
estimate the mixing time of $P$ given a common lower bound on most of
the entries in each column.
\begin{proposition}\label{dist-ub}
If $d(P) \leq \alpha$, then $\tau_{\mix} \leq \lceil
\log_{1/\alpha} 2e \rceil$.
\end{proposition}
%\begin{proof}
%From Aldous and Fill \cite{AF}: Let $\bar{d}(t) := \max_{x,x'}
%\frac{1}{2}||P^t(\cdot,x)-P^t(\cdot,x')||_1$ be the maximum pairwise
%column distance at time $t$.  This distance is submultiplicative;
%i.e., $\bar{d}(s+t) \leq \bar{d}(s)\bar{d}(t)$ for any $s,t \geq 0$.
%In particular, $\bar{d}(t) \leq \bar{d}(1)^t$, so for any $t \geq
%\lceil \log_{1/\alpha} 2e \rceil$ we have $\frac{1}{2}||P^t -
%\pi 1^\dagger||_1 \leq \bar{d}(\lceil \log_{1/\alpha} 2e \rceil) \leq
%\frac{1}{2e}$.\qed
%\end{proof}
\begin{proposition}\label{entry-lb}
If at least $\beta N$ entries in each column of $P$ are bounded below
by $\gamma / N$, where $\beta > \frac{1}{2}$ and $\gamma > 0$, then
$d(P) \leq 1 - \gamma (1 - 2(1-\beta))$.
\end{proposition}
%\begin{proof}
%Recall that for any two distributions $p,q$ we have
%$\frac{1}{2}||p-q||_1 = 1 - \sum_k \min\{p_k,q_k\}$.  It follows that
%$d(P) \leq 1-(1-2(1-\beta))N \cdot \gamma/N = 1 - \gamma (1 - 2
%(1-\beta))$.\qed
%\end{proof}

\subsection{Quantum walks}

Heretofore, a {\em quantum walk} is a pair $\langle U, \omega_T
\rangle$ where the {\em transition rule} $U$ is a unitary operator
acting on a finite-dimensional Hilbert space
and the {\em measurement rule} $\omega_T$ is
a $T$-parametrized family of probability density functions on
$[0,\infty)$ characterizing the (random) time at which a total
measurement is performed on the Hilbert space.\footnote{Though
repeatedly measured quantum walks generate mixed states, our analysis
is simplest without the introduction of density matrices.  We refer
the reader lacking sufficient background in quantum computing to the
excellent text of Kitaev, Shen, and Vyalyi \cite{KSV}.}

The unitary transition rule determines the orbit of a pure
quantum state, or {\em wavefunction} ($l_2$-normalized complex
vector), just as a Markov chain (or stochastic transition rule)
determines the orbit of a classical distribution ($l_1$-normalized
nonnegative vector).  Let $P$ be the Markov chain used in an MCMC
algorithm; in particular, $P$ is reversible.  Two natural
quantizations of $P$ are:
(a) the $\S \times \S$ unitary {\em continuous-time walk} given by
$U_{ct}(P) = \exp(iDPD^{-1})$ \cite{FG,AT}, where $H=DPD^{-1}$ is the
time-independent {\em Hamiltonian}, and (b) the $\S^2 \times
\S^2$ unitary discrete-time {\em Grover/Szegedy walk} given by
$U_{dt}(P) = (RS)^2$ \cite{Sze}, where $S$ is the involution
$\sum_{x,y \in \S} \ket{x,y} \mapsto \ket{y,x}$, $R$ is the reflection
$\sum_{x \in \S} \ketbra{x}{x} \otimes (2 \ketbra{p_x}{p_x} - I)$, and
$\ket{p_x}$ is the vector $\sum_{y \in \S} \sqrt{P(y,x)}\ket{y}$.
The quantization $U_{ct}(P)$ was used by Childs et al. \cite{CCD}
to solve in polynomial time a natural oracle problem for which no
classical polynomial-time algorithm exists.  The quantization
$U_{dt}(P)$ was used by Szegedy \cite{Sze} to prove a quadratic
quantum speedup for the hitting time of any symmetric Markov chain.

Measurement collapses the wavefunction to a classical distribution
according to the map $\ket{\phi} \mapsto \sum_x \ket{x}
|\braket{x}{\phi}|^2$.  For a quantum walk $\langle U, \omega_T
\rangle$, the Markov chain {\em generated by} the quantum walk is
given by $\hat{P}_T(y,x) := E_{t \leftarrow \omega_T}[|\braket{y}{U^t|x}|^2]$,
where $E$ denotes expected value.  We say that the quantum walk {\em
threshold-mixes} if the Markov chain it generates mixes in time
$O(1)$.  Examples of measurement rules from
the literature include the point distribution $\delta_T(t) :=
\delta(t-T)$ where $\delta$ is the delta function \cite{NV,ABNVW}, the
uniform distribution $\bar{\mu}_T := \frac{1}{T}\chi_{[0,T]}$ and its
discrete-time counterpart $\bar{\nu}_T := \frac{1}{T}\chi_{[0..T-1]}$
where $\chi$ is the characteristic function \cite{AAKV}, the
exponential distribution $\tilde{\mu}_T(t) := \frac{1}{T}\exp(-t/T)$,
and the geometric distribution $\tilde{\nu}_T(t) := \frac{1}{T}
(1-\frac{1}{T})^t$.  The exponential and geometric distributions are
{\em memoryless} and describe the interarrival time between
measurements in a Poisson process with measurements occurring at rate
$\lambda=1/T$ and a Bernoulli process with measurements occurring with
probability $p=1/T$ at each timestep, respectively.  These processes
coincide with the decoherence models of Alagic and Russell \cite{AR}
and Kendon and Tregenna \cite{KT}, respectively.

Nayak et al. \cite{NV,ABNVW} and Aharonov et al. \cite{AAKV} showed
that the so-called {\em Hadamard} walks $\langle U_{Had}, \delta_T
\rangle$ and $\langle U_{Had}, \bar{\nu}_T \rangle$ on the cycle $\Z_n$
threshold-mix quadratically faster than the classical random walk, although
their definitions of threshold mixing are slightly different than
ours.  Moore and Russell \cite{MR} showed that the continuous-time
walk $\langle U_{ct}(P(G)), \delta_T \rangle$ and the Grover walk
$\langle U_{dt}(P(G)), \delta_T \rangle$ on the hypercube $G=\Z_2^d$
mix perfectly and almost perfectly, respectively, in time $T=O(d)$.
They also showed that the continuous-time quantum walk on the hypercube with
measurement $\bar{\mu}_T$ does not mix to the uniform stationary
distribution $\pi$ of $P(\Z_2^d)$ in the limit $T \rightarrow \infty$.
Gerhardt and  Watrous \cite{GW} showed the same for a continuous-time
quantum walk on the symmetric group with measurement $\bar{\mu}_T$.
See the survey by Kempe \cite{Kem} for further results on quantum
walks.

We remark that there is a nice way to use the quantization $U_{dt}(P)$
to solve the mixing problem in time $O(1/\sqrt{\delta \pi_*})$ which, although
prohibitively costly in $\pi_*$, exhibits the desired dependence on
$\delta$.  Consider the stationary eigenstate $\ket{\tilde{\pi}} :=
\sum_{x \in \S} \sqrt{\pi_x} \ket{x}\ket{p_x}$ of $U_{dt}(P)$.  It is
clear that we can retrieve a good approximation to the classical
distribution $\pi$ by generating and then measuring a good
approximation to $\ket{\tilde{\pi}}$.\footnote{Similarly, we can
retrieve a good approximation to $\pi$ by generating and then
measuring a good approximation to the ground state $\ket{\pi} =
\sum_{x \in \S} \sqrt{\pi_x} \ket{x}$ of the Hamiltonian $H =
DPD^{-1}$ \cite{AT}.}
Magniez et al. \cite{MNRS} observe that $O(1/\sqrt{\delta})$ steps of
phase estimation on $U_{dt}(P)$ enable us to reflect about
$\ket{\tilde{\pi}}$.  By alternating this reflection with a reflection
about $\ket{\tilde{z}} := \ket{z}\ket{p_z}$ where $z$ is the initial
walk state (in particular, $\braket{\tilde{z}}{\tilde{\pi}} \geq
\sqrt{\pi_*}$), we can generate $\ket{\tilde{\pi}}$ from $\ket{\tilde{z}}$
in time $O(1/\sqrt{\delta \pi_*})$.  In fact, this algorithm is described
by Magniez et al. \cite{MNRS} as a {\em hitting} algorithm (i.e.,
generate the {\em unknown} state $\ket{\tilde{z}}$ from the fixed initial
state $\ket{\tilde{\pi}}$); the idea of running a quantum hitting
algorithm {\em in reverse} as a mixing algorithm was suggested by
Childs \cite{Chi}.

\section{Mixing properties of decoherent quantum walks}

\subsection{Two types of convergence}

Let $\hat{P}_T$ be the Markov chain generated by a quantum walk
$\langle U, \omega_T \rangle$.  Then repeating the quantum walk $T'$
times in succession generates the Markov chain $(\hat{P}_T)^{T'}$.
The following lemma (a variant of Theorem 3.4 in Aharonov et
al. \cite{AAKV}) and
theorem describe the asymptotic behavior of $\hat{P}_T$ and
$(\hat{P}_T)^{T'}$ in the limits $T \rightarrow \infty$ and $T'
\rightarrow \infty$, respectively.  For concreteness we will take
$U=U_{ct}$ in this subsection and the next; it is a simple exercise to
extend the results to discrete-time walk variants.  Although stated
explicitly for quantum walks, the results apply to any
time-independent quantum dynamics on a finite-dimensional Hilbert
space subjected to random destructive measurements.

\begin{lemma}[The limit $T \rightarrow \infty$]\label{single}
Let $P$ be a symmetric Markov chain and $\omega_T$ be a family of
distributions satisfying $E_{t \leftarrow \omega_T}[e^{i\theta t}]
\rightarrow 0$ as $T \rightarrow \infty$ for any $\theta \neq 0$.
In the limit $T \rightarrow \infty$, the Markov chain $\hat{P}_T$
generated by the quantum walk $\langle U_{ct}(P), \omega_T \rangle$
approaches the Markov chain $\Pi$ with entries
\begin{equation}\label{pimatrix}
\Pi(y,x) := \sum_j \bigl| \sum_{k \in C_j}
                   \braket{y}{\phi_k}\braket{\phi_k}{x} \bigr|^2
\end{equation}
where $\{\lambda_k,\ket{\phi_k}\}$ is the spectrum of $P$
and $\{C_j\}$ is the partition of these indices $k$ obtained by
grouping together the $k$ with identical $\lambda_k$.
\end{lemma}
\begin{proof}
Decomposing the quantum walk along spectral components gives us:
\begin{equation}
\hat{P}_T(y,x) = E_{t \leftarrow \omega_T}[|\sum_k \braket{y}{\phi_k}
  \braket{\phi_k}{x} e^{i \lambda_k t}|^2]
\end{equation}
Writing $| \cdot |^2$ as a product of complex conjugates, we obtain:
\begin{eqnarray}
\hat{P}_T(y,x)
 & = & E_{t \leftarrow \omega_T}[(\sum_k \braket{y}{\phi_k} \braket{\phi_k}{x})
 (\sum_l \braket{\phi_l}{y} \braket{x}{\phi_l}) e^{i(\lambda_k-\lambda_l)t}]\\
 & = & (\sum_k \braket{y}{\phi_k} \braket{\phi_k}{x})(\sum_l \braket{\phi_l}{y}
 \braket{x}{\phi_l}) E_{t \leftarrow \omega_T}[e^{i(\lambda_k-\lambda_l)t}]
\end{eqnarray}
Now by assumption, $E_{t \leftarrow \omega_T}[e^{i(\lambda_k-\lambda_l)t}]$
vanishes as $T \rightarrow \infty$ for all $\lambda_k \neq \lambda_l$,
so we have
\begin{equation}
\hat{P}_T(y,x)
 \rightarrow \sum_k \braket{y}{\phi_k} \braket{\phi_k}{x}
 (\sum_{l : \theta_l = \theta_k} \braket{\phi_k}{y} \braket{x}{\phi_k})
 = \sum_j |\sum_{k \in C_j} \braket{y}{\phi_k} \braket{\phi_k}{x}|^2
 = \Pi(y,x)
\end{equation}
in the limit $T \rightarrow \infty$.\qed
\end{proof}

It can be inferred from this lemma that most quantum walks converge to
a distribution $\rho$ other than the uniform stationary distribution
$\pi = u$, and that $\rho$ is not even independent of the initial walk
state.\footnote{For a walk on the symmetric group $S_n$, Gerhardt
and Watrous \cite{GW} showed that $\frac{1}{2}||\Pi - u 1^\dagger||_1
\geq \frac{1}{n}-\frac{1}{n \cdot n!}\binom{2n-2}{n-1}$; for a walk on
the hypercube $\Z_2^n$, Moore and Russell \cite{MR} showed that there
exists an $\epsilon > 0$ such that $\frac{1}{2}||\Pi - u 1^\dagger||_1
\geq \epsilon$.}  There are exceptions to this rule, for example
quantum walks with distinct eigenvalues on Cayley graphs of Abelian
groups (as observed by Aharonov et al. \cite{AAKV}), but they are not
likely to arise in MCMC applications where the Markov chains have
little structure.  How then are we to sample from $u$ using
quantizations of these Markov chains?  Here is where decoherence helps.

\begin{theorem}[The limit $T' \rightarrow \infty$]\label{repeated}
Let $P$ be a symmetric, irreducible Markov chain and $\omega_T$ be a
family of distributions satisfying $E_{t \leftarrow \omega_T}[e^{i\theta t}] 
\rightarrow 0$ as $T \rightarrow \infty$ for any $\theta \neq 0$.  For $T$
sufficiently large (but fixed), the $T'$-repeated quantum walk
$\langle U_{ct}(P), \omega_T \rangle$ generates a Markov chain
$(\hat{P}_T)^{T'}$ approaching $u 1^\dagger$ in the limit $T'
\rightarrow \infty$.
\end{theorem}
\begin{proof}
We need to show that for $T$ sufficiently large, the Markov chain
$\hat{P}_T$ is ergodic with uniform stationary distribution.

That the uniform distribution is stationary is clear:  each of the
$P_t(y,x) := |\braket{y}{e^{iPt}|x}|^2$ has uniform stationary
distribution since the uniform classical state is invariant
under unitary quantum operations and under total measurement of the
system; thus, any probabilistic combination $\hat{P}_T$ of them has
uniform stationary distribution.

To show that $\hat{P}_T$ is ergodic for all sufficiently large
$T$, it is sufficient (by Lemma \ref{single}) to prove that $\Pi$ is
ergodic.  (The latter implies the former because the ergodic
matrices form an open subset of the set of stochastic matrices.)  Why
is $\Pi$ ergodic?  Because the $1$-eigenspace of $P$ is precisely the
space spanned by $u$, so it follows from Lemma \ref{single} (by
consideration of only this nondegenerate eigenspace in the expression
(\ref{pimatrix})) that $\Pi(y,x) \geq 1/N^2$ for every $x,y$.

In fact, each of the $P_t$ (and so $\hat{P}_T$ and $\Pi$ as well) is
symmetric.\footnote{$\Pi$ is also positive semidefinite: it is the
Gram matrix of $\{f_s\}$ with
$f_s(kl) := \braket{s}{\phi_k} \braket{\phi_l}{s}$ if $\lambda_k =
\lambda_l$, $0$ otherwise.}  To see this, write out the Taylor series
for $\exp(iPt)$ and note that every positive integer power $P^k$ is
symmetric (since $P^2(x,y) = \sum_z P(x,z) \cdot P(z,y) = \sum_z
P(y,z) \cdot P(z,x) = P^2(y,x)$).  This property will be quite useful
in the next subsection: it will allow us to use Theorem \ref{ds-ineq}
to relate the spectral gap and the mixing time of $\hat{P}_T$.\qed
\end{proof}

\subsection{Invariance of the mixing time}

Consider the quantum walks $\langle U_{ct}(P), \bar{\mu}_T \rangle$
and $\langle U_{ct}(P), \tilde{\mu}_T \rangle$ where $P$ is a
symmetric Markov chain.  We show that these two quantum walks mix with
essentially the same efficiency.  The result extends beyond
$\bar{\mu}, \tilde{\mu}$ to any pair of measurements $\omega, \omega'$
which are sufficiently smooth or have nontrivial overlap as
distributions.

Let $\bar{P}_T$ and $\tilde{P}_T$ be the Markov chains generated by
the walks with measurement $\bar{\mu}$ and $\tilde{\mu}$,
respectively, and let $\bar{\delta}_T := 1 - ||\bar{P}_T |_{u^\perp}
||_2$ and $\tilde{\delta}_T := 1 - ||\tilde{P}_T |_{u^\perp} ||_2$ be
their respective spectral gaps.

\begin{lemma}[Spectral gap inequality]\label{gap}
Let $\bar{\delta}_T$ and $\tilde{\delta}_T$ be defined as above.  Then
for any $k \geq 1$ we have the inequality:
\begin{equation}
e^{-1}\bar{\delta}_T \leq \tilde{\delta}_T \leq k(1-e^{-k}) \cdot
\bar{\delta}_{kT} + 2e^{-k}
\end{equation}
\end{lemma}
\begin{proof}
Suppose we want to simulate $\bar{P}_T$ by $\tilde{P}_T$.  Scaling
the distribution $\bar{\mu}_T$ by $\alpha := 1/e$ allows us to ``fit
it inside'' the distribution $\tilde{\mu}_T$ (i.e., $e^{-1}
\bar{\mu}_T \leq \tilde{\mu}_T$ pointwise), so we can express
$\tilde{\mu}_T$ as the probabilistic combination $\alpha \bar{\mu}_T +
(1-\alpha)\nu$ for some distribution $\nu$, so that
\begin{equation}
\tilde{P}_T = E_{t \leftarrow \tilde{\mu}_T}[P_t] = \alpha E_{t
  \leftarrow \bar{\mu}_T}[P_t] + (1-\alpha) E_{t \leftarrow \nu}[P_t]
  = \alpha \bar{P}_T + (1-\alpha) Q
\end{equation}
where $Q$ is stochastic with uniform stationary distribution.  It
follows that
\begin{equation}
||\tilde{P}_T |_{u^\perp} ||_2 \leq 1/e ||\bar{P}_T |_{u^\perp}||_2 +
(1-1/e) ||Q |_{u^\perp}||_2
\end{equation}
which implies that $\tilde{\delta}_T \geq 1/e \cdot \bar{\delta}_T$
since $||Q |_{u^\perp}||_2 \leq 1$.

Suppose we want to simulate $\tilde{P}_T$ by $\bar{P}_{kT}$.  Then the
basic approach is the same, but since the support of $\tilde{\mu}_T$ is
not compact we have to be careful.  Scaling the distribution
$\tilde{\mu}_T$ by $\beta := 1/k$ allows us to fit it inside the distribution
$\bar{\mu}_{kT}$ up to the point $t = kT$, and the probability mass in
$\tilde{\mu}_T$ past $t = kT$ is only $\Pr_{t \leftarrow
\tilde{\mu}_T}[t > kT] = e^{-k}$.  So we can write
\begin{equation}
\tilde{\mu}_T = (1-e^{-k}) \cdot \tilde{\mu}_T^{head} + e^{-k} \cdot
\tilde{\mu}_T^{tail}
\end{equation}
where $\tilde{\mu}_T^{head}$ and $\tilde{\mu}_T^{tail}$ are the
conditional distributions of $\tilde{\mu}_T$ such that $t \leq kT$ and
$t > kT$, respectively; thus,
\begin{equation}
\tilde{P}_T = (1-e^{-k}) \cdot \tilde{P}_T^{head} + e^{-k} \cdot
\tilde{P}_T^{tail}
\end{equation}
where $\tilde{P}_T^{head}$ and $\tilde{P}_T^{tail}$ are the
expectations of $P_t$ with respect to $\tilde{\mu}_T^{head}$ and
$\tilde{\mu}_T^{tail}$, respectively.  Since we can fit
$\tilde{\mu}_T^{head}$ inside $\bar{\mu}_{kT}$ if we scale it by
$1/k$, we can write
\begin{equation}
\bar{P}_{kT} = \frac{1}{k} \tilde{P}_T^{head} + (1-\frac{1}{k})Q
\end{equation}
where $Q$ is stochastic with uniform stationary distribution.  The
above equations yield:
\begin{eqnarray}
\bar{P}_{kT} & = & \frac{1}{k(1-e^{-k})} (\tilde{P}_T - e^{-k}
                   \tilde{P}_T^{tail}) + (1-\frac{1}{k})Q \nonumber\\
             & = & \frac{1}{k(1-e^{-k})} \tilde{P}_T -
                   \frac{e^{-k}}{k(1-e^{-k})} \tilde{P}_T^{tail} +
                   (1-\frac{1}{k}) Q
\end{eqnarray}
From the triangle inequality, we obtain
\begin{equation}
||\bar{P}_{kT} |_{u^\perp}||_2 \leq \frac{1}{k(1-e^{-k})}
  ||\tilde{P}_T |_{u^\perp}||_2 + \frac{e^{-k}}{k(1-e^{-k})}
  ||\tilde{P}_T^{tail} |_{u^\perp}||_2 + (1-\frac{1}{k})||Q |_{u^\perp}||_2
\end{equation}
and, rearranging terms and simplifying:
\begin{equation}
\frac{1}{k(1-e^{-k})} (1 - ||\tilde{P}_T |_{u^\perp}||_2) -
\frac{2e^{-k}}{k(1-e^{-k})} \leq 1 - ||\bar{P}_{kT} |_{u^\perp}||_2
\end{equation}
\qed
\end{proof}

\begin{theorem}[Equivalence of measurements]\label{equivalence}
Let $P$ be a symmetric Markov chain.  Then: (a) if the $T'$-repeated
quantum walk $\langle U_{ct}(P), \bar{\mu}_T \rangle$ threshold-mixes,
then the $T' \cdot O(\log N)$-repeated quantum walk $\langle
U_{ct}(P), \tilde{\mu}_T \rangle$ threshold-mixes; (b) if the
$T'$-repeated quantum walk $\langle U_{ct}(P), \tilde{\mu}_T \rangle$
threshold-mixes, then the $T' \cdot O(\log T' \log N)$-repeated quantum
walk $\langle U_{ct}(P), \bar{\mu}_{T \cdot O(\log T')} \rangle$
threshold-mixes.
\end{theorem}
\begin{proof}
To see (a), note that our assumption implies that $\bar{P}_T$ mixes in
time $T'$.
%Therefore, $\bar{\delta}_T = \Omega(\lambda / T')$ by
%Theorem \ref{ds-ineq}, where $\lambda = 1 - \bar{\delta}_T$.  We will
%assume that $\lambda$ is bounded away from $0$; if it were not, then
%$\bar{\delta}_T$ would be, and the quantum walk would be unnecessary
%because the classical walk would already mix optimally fast.  Then
%$\bar{\delta}_T = \Omega(1 / T')$,
Therefore, $\bar{\delta}_T = \Omega(1/T')$ by Theorem \ref{ds-ineq},
and from Lemma \ref{gap} it follows that $\tilde{\delta}_T = \Omega(1
/ T')$.  Applying Theorem \ref{ds-ineq} again, we obtain for
$\tilde{P}_T$ a mixing time of $O(T' \log N)$.

The proof of (b) is almost as straightforward.  Our assumption implies
that $\tilde{P}_T$ mixes in time $T'$, so $\tilde{\delta}_T = \Omega(1
/ T')$ by Theorem \ref{ds-ineq}.  Set $k$ to be the smallest integer
for which $\tilde{\delta}_T \geq 3e^{-k}$; in particular, $k =
\Theta(\log \tilde{\delta}_T^{-1}) = O(\log T')$.  By Lemma \ref{gap}:
\begin{equation}
\bar{\delta}_{kT} \geq \frac{1}{k(1-e^{-k})} (\tilde{\delta}_T -
2e^{-k}) \geq \frac{1}{k(1-e^{-k})} (e^{-k}) =
\Theta(\frac{\tilde{\delta}_T}{\log \tilde{\delta}_T^{-1}}) =
\Theta(\frac{1}{T' \log T'})
\end{equation}
Applying Theorem \ref{ds-ineq} again, we obtain for $\bar{P}_{kT}$ a
mixing time of $O(T' \log T' \log N)$.\qed
\end{proof}

It should be readily apparent that this equivalence holds for
any two measurement rules with finite expectation and significant
overlap for most $T$.  We also remark that although the above lemma
and theorem are stated in terms of quantum walks, the proofs indicate
that they are merely statements about an abstract game involving a
collection of symmetric Markov chains $\{P_t\}_{t \geq 0}$ and
a $T$-parametrized family of probability measures $\{\omega_T\}$,
where we seek to minimize the ``cost function'' $T \cdot T'$.

\section{Quantum speedup for periodic lattices}

\subsection{A near-diameter upper bound}

The classical random walk on the periodic lattice $\Z_n^d$ (with $N =
n^d$ vertices and diameter $\lfloor n/2 \rfloor d$) has uniform stationary
distribution $\pi = u$ and spectral gap $\delta = \Theta(\min \{
\frac{1}{d}, \frac{1}{n^2} \})$.  It threshold-mixes in time
$\Theta(n^2 d \log d)$, which is $O(\sqrt{\delta^{-1}} \log 1/\pi_*)$
only when $\Z_n^d$ is quite high-dimensional: in particular, when $d$
is roughly of order $n^4$ or larger.  We show that a few repetitions
of the continuous-time quantum walk can bring this down to $O(n d \log
d)$, which is $O(\sqrt{\delta^{-1}} \log 1/\pi_*)$ for any $d \geq 1$ and
$n \geq 2$ and asymptotically no worse than the diameter of $\Z_n^d$
up to at most a logarithmic factor.

First we prove a lemma governing mixing of various decoherent quantum
walks on the cycle $\Z_n$.

\begin{lemma}[Mixing on cycles]\label{cycles}
Let $\Z_n$ be the cycle on $n \geq 2$ vertices.  The continuous-time
walks $\langle U_{ct}(P(\Z_n)),\omega_T \rangle$ with measurement $\omega
\in \{\delta, \bar{\mu}, \tilde{\mu}\}$ threshold-mix for any $T \in
\I := [\frac{2}{3}\cdot\frac{n}{2},\frac{n}{2}]$, and the Hadamard
walks $\langle U_{Had}(\Z_n),\omega_T \rangle$ with
measurement $\omega \in \{\bar{\nu}, \tilde{\nu}\}$ threshold-mix for
any $T \in \J := [\frac{2}{3}\cdot\frac{n}{\sqrt{2}},\frac{n}{\sqrt{2}}]$.
\end{lemma}
\begin{proof}
Consider first the continuous-time walk.  To prove that it
threshold-mixes with any of the measurements $\omega \in \{\delta, \bar{\mu},
\tilde{\mu}\}$ for any $T \in \I$, it suffices by Proposition \ref{dist-ub} to
show that for every $t \in \I' := [\frac{3}{5}\cdot\frac{n}{2},
\frac{4}{5}\cdot\frac{n}{2}]$, $d(P_t)$ is  bounded below one by a
positive constant, where $P_t$ is the Markov chain generated with
measurement $\omega = \delta$.  (Indeed, this easily implies that
$d(\bar{P}_T)$ and $d(\tilde{P}_T)$ are bounded below one by a smaller
positive constant, where $\bar{P}_T$ and $\tilde{P}_T$ are the Markov
chains generated with measurement $\omega = \bar{\mu}$ and $\omega =
\tilde{\mu}$, respectively.)

Let $\ket{\phi_t}$ and $\ket{\psi_t}$ be the wavefunctions at time
$t$ for the continuous-time walks on $\Z$ and $\Z_n$, respectively,
starting from the origin (without loss of generality, since $\Z$ and
$\Z_n$ are vertex-transitive).  Then for each $\bar{y} \in \Z_n$ we have:
\begin{equation}
\braket{\bar{y}}{\psi_t}
 = \sum_{y \equiv \bar{y} \bmod n} \braket{y}{\phi_t}
\end{equation}
Childs \cite{Chi} shows that $\braket{y}{\phi_t} = (-i)^{y}
J_{y}(t)$, where $J_{y}$ is a Bessel function of the first kind.
In particular, for $|y| \gg 1$ the quantity $|J_{y}(t)|$ is (a)
exponentially small in $|y|$ for $t < (1-\epsilon) \cdot |y|$ and (b)
of order $|y|^{-1/2}$ for $t > (1+\epsilon) \cdot |y|$.  For every $t
< \frac{4}{5} \cdot \frac{n}{2}$, property (a) implies that the only
term in the above summand that is non-negligible is the
$\braket{y}{\phi_t}$ with $|y| < \frac{n}{2}$ (call it $\hat{y}$
and note that $\bar{y} \leftrightarrow \hat{y}$ is a 1-1
correspondence), so we can use property (b) to conclude that up to
a negligible correction
\begin{equation}
|\braket{\bar{y}}{\psi_t}| \approx
  |\braket{\hat{y}}{\phi_t}| = \Theta(1/\sqrt{n})
\end{equation}
for every $|\hat{y}| \gg 1$ and $t > (1+\epsilon) \cdot |\hat{y}|$.
In particular, the nearly $\frac{3}{5}n$ different $\bar{y}$ with $1
\ll |\hat{y}| \leq \frac{3}{5} \cdot \frac{n}{2}$ satisfy
$|\braket{\bar{y}}{\psi_t}| = \Omega(1/\sqrt{n})$, and therefore
$P_t(\bar{y},\bar{0}) = \Omega(1/n)$, for every $t \in \I'$.  So by
Proposition \ref{entry-lb}, $d(P_t)$ is bounded below one by a
positive constant.

For the Hadamard walk, the wavefunction is no longer characterized by
Bessel functions, but it retains the same essential asymptotic
spreading behavior as its continuous-time counterpart (see Nayak et
al. \cite{NV,ABNVW}), and the argument above works with little
modification.  A caveat is the emergence of a parity problem: if $n$
is even, then $\Z_n$ is bipartite and the wavefunction is supported
only on vertices of the same parity at each integer timestep.  Hence
the Hadamard walk with $\omega = \delta$ threshold-mixes only on
vertices of the same parity, but with time-averaged measurement $\omega
= \bar{\nu}$ or $\omega = \tilde{\nu}$ parity is broken and
threshold-mixing occurs on the entire vertex set.

Although the argument above relies on the asymptotic behavior of the
wavefunction as $n \rightarrow \infty$, this is clearly the difficult
case: if $n$ is bounded, then it suffices to show only that there exists a
time (or a pair of consecutive timesteps, in the case of the Hadamard
walk) in which the wavefunction is supported on at least $2/3$ of the
vertices.\qed
\end{proof}

For the Hadamard walk with measurement $\omega = \tilde{\nu}$, Lemma
\ref{cycles} resolves a conjecture of Kendon and Tregenna \cite{KT}
based on numerical experiments.

Let $G^d$ denote the $d$th (Cartesian) power of a graph $G$.  Examples
are the $d$-dimensional standard lattice (the $d$th power of a line)
and the $d$-dimensional periodic lattice (the $d$th power of a
cycle).  The following theorem shows how to extend a threshold-mixing
result from $G$ to $G^d$.

\begin{theorem}[Mixing on graph powers]\label{powers}
Suppose the continuous-time quantum walk $\langle U_{ct}(P(G)),\delta_T
\rangle$ threshold-mixes.  Then the $O(\log d)$-repeated walk
$\langle U_{ct}(P(G^d)),\delta_{Td} \rangle$ threshold-mixes.
\end{theorem}
\begin{proof}
The Hamiltonian $H' = P(G^d)$ is related to the Hamiltonian $H =
P(G)$ by the identity:
\begin{equation}
H' = \frac{1}{d} \sum_{j=1}^d I^{\otimes (j-1)} \otimes H \otimes
                       I^{\otimes (d-j)}
\end{equation}
Since $H'$ commutes with the identity $I$, which can introduce at
most a global phase factor to the system, the Markov chain $P'_t$
generated by the walk $\langle U_{ct}(P(G^d)),\delta_t \rangle$ is the
$d$th tensor power of the Markov chain $P_{t/d}$ generated by the walk
$\langle U_{ct}(P(G)),\delta_{t/d} \rangle$.
%the distribution induced on
%the vertices of $G^d$ by $H'$ at time $t$ is the $d$th tensor power of
%the distribution induced on the vertices of $G$ at time $t/d$.
By assumption, $d(P_T) \leq \alpha$ for some constant $\alpha < 1$.
By Proposition \ref{dist-ub}, we can choose $T' = O(\log d)$ to ensure
that:
\begin{equation}
\frac{1}{2}||(P_T)^{T'} - u 1^\dagger||_1 \leq d((P_T)^{T'}) \leq
\frac{1}{6d^2}
\end{equation}
%(mynote: $\ln (3/2) \mapsto \frac{1}{3}$ heretofore)
Then at least $n\sqrt[d]{2/3}$ entries in each column of $(P_T)^{T'}$
are bounded below by $\frac{1-1/2d}{n}$, otherwise we would have the
contradiction
\begin{equation}
\frac{1}{2} ||(P_T)^{T'} - u 1^\dagger||_1
 = 1 - \sum_y \min \{ (P_T)^{T'}(y,x), \frac{1}{n} \}
 > (1-\sqrt[d]{2/3}) \frac{1}{2d}
 \geq \frac{1}{6d^2}
\end{equation}
where the first equation uses the identity $\frac{1}{2}||p-q||_1 = 1 -
\sum_k \min\{p_k,q_k\}$ for distributions $p, q$ and the last
inequality uses simple algebra along with the fact that for any $d \geq 1$:
\begin{equation}
(1-\frac{1/3}{d})^d \geq \frac{2}{3}
\end{equation}
Since $(P'_{Td})^{T'} = ((P_T)^{T'})^{\otimes d}$, at least
$(n\sqrt[d]{2/3})^d = \frac{2}{3}n^d$ of the entries in each column of
$(P'_{Td})^{T'}$ are bounded below by $(\frac{1-1/2d}{n})^d \geq
\frac{1}{2 n^d}$.  It follows from Proposition \ref{entry-lb} that
$(P'_{Td})^{T'}$ threshold-mixes in time $O(1)$.\qed
\end{proof}

We have the following corollary for the $d$th power $\Z_n^d$ of the
cycle $\Z_n$.

\begin{corollary}[Mixing on periodic lattices]\label{lattices}
Let $\Z_n^d$ be the $d$-dimensional periodic lattice with $n \geq 2$
vertices per side.  The $O(\log d)$-repeated continuous-time quantum
walk $\langle U_{ct}(P(\Z_n^d)), \omega_{nd/2} \rangle$ with measurement
$\omega \in \{\delta, \bar{\mu}, \tilde{\mu}\}$ threshold-mixes.
\end{corollary}
\begin{proof}
Combining Lemma \ref{cycles} with Theorem \ref{powers}, we conclude
that $\langle U_{ct}(P(\Z_n^d)), \delta_T \rangle$ threshold-mixes for
any $T \in [\frac{2}{3}\cdot\frac{nd}{2}, \frac{nd}{2}]$.  It is easy
to see (cf. Lemma \ref{cycles}) that this is sufficient to imply the
stated corollary not only for the measurement $\omega = \delta$ but
also for the time-averaged measurements $\bar{\mu}$ and
$\tilde{\mu}$.\qed
\end{proof}

For $d \geq 1$, this extends the results of Fedichkin et
al. \cite{FST,SF1,SF2} by confirming $O(n)$ and $O(d \log d)$ scaling
(suggested by their analytical estimates and numerical experiments
in regimes of both high and low decoherence) of the fastest-mixing
walk, which they conjectured to be decoherent rather than unitary. 

\subsection{The Grover walk}

An important question is whether there is a $T'$-repeated Grover walk
$\langle U_{dt}(P(\Z_n^d)), \omega_T \rangle$ that threshold-mixes
for $T = O(nd)$ and $T' = O(\log d)$.  Szegedy \cite{Sze} showed that
the {\em phase gap} (minimum nonzero eigenvalue phase from
$[-\pi,\pi]$ in absolute value) of $U_{dt}(P)$
is $\Omega(\sqrt{\delta})$ and exploited this property to prove a
quadratic quantum speedup for the {\em hitting time} of any symmetric
Markov chain $P$.  A natural adaptation of his argument to the {\em
mixing time} setting would be something like the following: since the
phase gap of $U_{dt}(P)$ is $\theta = \Omega(\sqrt{\delta})$, we
expect to see by decomposing the action of $U_{dt}(P)$ along spectral
components that roughly $O(1/\theta) = O(\sqrt{\delta^{-1}})$ timesteps
suffice for the orbit of any initial basis state to ``cover'' the
entire state space with sufficient amplitude.

Unfortunately, this argument is incorrect: in fact, the orbit may
remain quite localized around the initial basis state.  This happens
with dramatic effect to the Grover walk on the complete graph
$G=K_N$, which mixes in time $T'=\Theta(N)$ (cf. \cite{Ric}) even
though the classical random walk on $K_N$ mixes in a single timestep.
It also happens to the Grover walk on $\Z_n^d$ for $d=2$, albeit less
dramatically \cite{IKK,TFMK,MBSS}.\footnote{For $d=1$ it does not; the
Grover walk on $\Z_n$ with measurement $\omega = \bar{\nu}$ mixes
perfectly in $T=n$ timesteps.}  The probability distribution $p_t$
induced on the vertices of $\Z_n^2$ at time $t \leq n/2$ is primarily
localized at the initial basis state \cite{IKK} but has substantial
secondary spikes which propogate across the lattice in orthogonal
directions \cite{MBSS}.  In particular, the standard
deviation of $p_t$ appears to grow linearly with $t$ \cite{MBSS} and
the mixing time of the Grover walk on $\Z_n^2$ with measurement
$\omega = \bar{\nu}$ appears to be $T=O(n)$.  In the high-dimensional
regime, Moore and Russell \cite{MR} proved that the Grover walk on
$\Z_2^d$ with measurement $\omega = \delta$ mixes almost perfectly in
time $T=O(d)$.  It seems plausible that the decoherent Grover walk on
$\Z_n^d$ with measurement $\omega = \bar{\nu}$ mixes as fast
asymptotically as the continuous-time walk.

%Examples: 1. Expanders ($\delta = \Theta(1)$) already mix optimally
%fast by a classical walk.  2. Johnson graph $J(n,r)$ on $\binom{n}{m}$
%$r$-subsets of an $n$-set ($\delta = 1/r$, $\log N = \log
%\binom{n}{m}$) mixes classically in time $O(r \cdot (1-\frac{r}{n})
%\log n)$ (see [DS]), which is apparently already $O(1/\sqrt{\delta}
%\log N)$ (but check this).  (So the quantum walk is not needed for
%$J(n,r)$, but note that it may be interesting nonetheless as it is
%periodic with period $2 \pi r (n-r)$ so we can use Matlab to plot its
%limiting distribution, which may be very lumpy due to eigenvalue
%degeneracies.. show that there's not catastrophic localization though?
%The eigenvalues of $J(n,r)$ are $1 - \frac{j(n-j+1)}{r(n-r)}$ with
%multiplicity $\binom{n}{j}-\binom{n}{j-1}$ for $j=1,\ldots,k$ and $1$
%with multiplicity $1$.)

\section{Conclusion and open problems}

We have shown that decoherent quantum walks have the potential to
speed up a large class of classical MCMC mixing processes.  Exactly
how large this class is, and whether a generic quantum mixing speedup
to $O(\sqrt{\delta^{-1}} \log 1/\pi_*)$ is possible, remain important
open questions.  Since it seems that quantum walks can outperform
classical random walks in low-dimensional examples and underperform in
very high-dimensional examples (such as the complete graph), a hybrid
method may work best for generic Markov chains consisting of both low-
and high-dimensional substructures.  Also worth investigating is
whether by randomizing the ``coin'' used in discrete-time quantum
walks we can improve mixing on the complete graph and other
adversarial examples.

%In 'our contributions', add: \footnote{Lest the reader find the
%notion of a {\em continuous-time} quantum walk unsettling, we remark
%that there are techniques available for simulating $U_{ct}$
%efficiently using standard quantum circuits \cite{Chi}.}

{\bf Acknowledgements.}  I would like to thank Mario Szegedy, Viv
Kendon, Fr\'{e}d\'{e}ric Magniez, Miklos Santha, Iordanis Kerenidis,
and Julia Kempe for useful discussions on quantum walks and Todd Brun
for helpful comments on the presentation.

%\newpage

\bibliographystyle{plain}

\end{document}